**ASTRONOMY, STATISTICS IN**

Author: Eric D. Feigelson

Astronomy is the science devoted to the study of planets, stars, galaxies and the Universe as a whole.  It is closely coupled to astrophysics where laws governing physical processes (e.g., gravity, electromagnetism, quantum mechanics) are used to model observed astronomical properties.  Perhaps more than other physical sciences, astronomy is frequently statistical in nature.  The objects under study are inaccessible to direct manipulation in the laboratory, so the astronomer is restricted to observing a few external characteristics and inferring underlying properties and physics.

As the oldest physical science, astronomy has served the development of statistical methodology for centuries.  During the 17-18$^{th}$ centuries, in particular, important foundations of modern statistical theory were formulated to address astronomical problems; the astronomers were the statisticians.  The fields diverged during much of the 20$^{th}$ century, as astronomy turned to physics for insights and statisticians turned to applications in human sciences and industries.  In the past two decades, the cross-disciplinary field of astrostatistics has reemerged to deal both with important astrophysical issues and to treat mega-datasets produced by high-technology observatories.

## HISTORY OF ASTROSTATISTICS

Quantitative measurements of celestial phenomena were carried out by many ancient civilizations (1,2,3,4). The classical Greeks were not active observers but were unusually creative in the applications of mathematical principles to astronomy. Greek natural philosopher Hipparchus started a discussion which lasted for centuries on how inconsistent measurements should be combined to quantify a phenomenon. Finding scatter in Babylonian measurements of the length of a year, defined as the time between solstices, he took the middle of the range -- rather than the mean or median -- for the best value.  Ptolemy similarly estimated parameters of a non-linear cosmological model using a minimax goodness-of-fit method.  The Persian 11$^{th}$ century scientist Al-Biruni recommended a three-point procedure, and discussed the dangers of propagating errors from inaccurate instruments and inattentive observers.  While some medieval scholars advised against the acquisition of repeated measurements, fearing that errors would compound rather than compensate for each other, the utility of the mean to increase precision was demonstrated with great success in the 16$^{th}$ century by Tycho Brahe.

Celestial mechanics in the eighteenth century, in which Newton's law of gravity was found to explain even the subtlest motions of heavenly bodies, required the derivation of a few interesting quantities from numerous inaccurate observations. This required advances in the understanding of statistical inference and error distributions. Mayer, in his 1750 study of lunar librations, suggested a procedure of reconciling a system of 27 inconsistent linear equations in three unknowns by solving the equations in groups. Laplace, in a 1787 analysis of the influence of Jupiter's gravity on Saturn's motion, suggested a more unified approach that led to Legendre's invention of the least-squares method in an 1805 study of cometary orbits. Shortly thereafter, in an 1809 monograph on the mathematics of planetary orbits, Gauss first presented the normal (or Gaussian) distribution of errors in overdetermined systems of equations using a form of Bayes' theorem, though the actual derivation was flawed. Legendre developed $L_2$ least squares parameter estimation to model cometary orbits. The least-squares method rapidly became widely used in European astronomy and geodesy.

Many other individuals also contributed substantially to both astronomy and statistics
Galileo proposed an $L_1$ procedure for estimation and discussed observational errors concerning the distance to the supernova of 1572. Huygens wrote books on optics and Saturn and on probability in games of chance. Halley, famous for his early contributions to celestial mechanics, laid important foundations to mathematical demography and actuarial science. Bessel, codiscoverer of stellar parallaxes and the binary companion of Sirius, introduce the notion of probable error. Quetelet, founder of the Belgian Royal Observatory, led early applications of probability theory to the social sciences. Airy, a late-19$^{th}$ century British Royal astronomer, is known both for his text on statistical errors and his study of telescope optics.

Relatively few connections between astronomy and statistics are notable in the first decades of the 20$^{th}$ century. A subfield called ``statistical astronomy" concentrated on issues involving positional astronomy, star counts and Galactic structure (5). Astronomers were mainly familiar with normal distributions and least-squares estimation methods as presented in volumes written for physical scientists, particularly (6). Fourier analysis commonly used for time series analysis. The 1970s witnesses some broadening of methodology as maximum likelihood estimation and the nonparametric Kolmogorov-Smirnov goodness-of-fit test gained popularity. Bayesian methods have become increasingly important since the 1990s.

The modern field of astrostatistics grew in the 1990s, stimulated by monographs on statistical aspects of astronomical image processing (7,8), galaxy clustering (9), Bayesian analyses (10) and astrostatistics in general (11).  The continuing conference series "Statistical Challenges in Modern Astronomy" (12,13) brought together astronomers and statisticians.   Collaborations between astronomers and statisticians emerged such as the California/Harvard AstroStatistical Collaboration (14).   The education of astronomers in statistical methodology remains weak.  A Center for Astrostatistics has operated recently to run a methodological summer school for astronomers, provide computer codes, and otherwise liaise the communities (15).   A selection of modern astrostatistical issues is reviewed in the remainder of this article.

**ASTRONOMICAL MEGA-SURVEYS AND MULTIVARIATE ANALYSIS**

Resources for astronomical measurements are increasingly focused on well-designed large-scale surveys.  These entail imaging large areas of the sky at various wavebands and/or obtaining spectra of large number of objects.  Perhaps the most influential recent project has been the Sloan Digital Sky Survey (16) which is producing terabytes of images, catalogs of ~200 million celestial objects, and spectra of ~1 million galaxies, quasars and stars.  Several hundred publications have emerged since it began in 2000 on a wide range of astrophysical issues.  Data from numerous surveys are freely available to the worldwide research community through the Virtual Observatory project (17).

A typical astronomical catalog is a table where rows represent individual objects and columns denote measurements of various properties: location in the sky (astrometry), brightness at different wavebands of light (photometry), morphological and categorical information.   The structure of such tables can be fruitfully studied with well-established techniques of multivariate analysis and classification.  However, astronomical multivariate databases differ from those obtain in other fields due to three interrelated characteristics: heteroscedastic measurement errors, censoring and truncation survey (18).  Based on careful calibration of the instruments and observation of source-free regions of the sky, astronomers directly measure the uncertainty of each photometric measurement.  Thus, each brightness value is accompanied by a heteroscedastic measurement error with known variance.  Analysis of the database thus needs multivariate methods with heteroscedastic weightings, which are generally not available.  It is also common that some properties of a source are too faint to be detected, leading to left-censoring of the brightness based on the known measurement error (19).  Astronomy thus needs extensions to survival analysis that permit censoring in any variable.  Finally, the list of objects is often truncated by the sensitivity limits of the underlying.

Steps have been made in addressing these statistical challenges.  Astronomer Lynden-Bell (20) and statistician Woodroofe  (21) derived the nonparametric maximum-likelihood estimator for a randomly truncated sample.  Other results include: a 2-sample test for randomly truncate datasets (22), an extension to Kendall's tau correlation with censoring for a 3-variable database (23), a test for completeness in a truncated sample (24), a semiparametric estimator for an irregularly truncated sample (25), and a thorough Bayesian treatment of linear regression including heteroscedastic measurement error, censoring and truncation (26).

## EXTRASOLAR PLANETS AND TIME SERIES ANALYSIS

Despite the apparent constancy of celestial objects to the human eye, they exhibit an incredible array of variable phenomena when studied with modern telescopes (27, 28). Stellar rotations, pulsations, and orbits of binary stars and planetary systems produce periodic variations typically studied with Fourier analysis. Distant blazers produce aperiodic correlated variations. Accretion onto compact objects, including black holes, produces combinations of periodic and stochastic variations. Solar flares, supernovae, and gamma-ray bursts produce short-lived explosive variations. Variability is seen at radio, infrared, visible, X-ray and gamma-ray wavebands. Analysis is often hindered by unevenly-spaced and heteroscedastically weighted data points. An important method is a generalization of Schuster's periodogram for unevenly-space data (29).

The analysis of binary star and planetary orbits has its roots in the analyses of Laplace based on Newtonian gravity. A binary star system follows an elliptical orbit but the modeling of its physical properties can be complex due to orbital eccentricity, inclination to the line of sight, tidal distortions, reflection of the other star's light, and so forth. Binary star models, sometimes with dozens of nonlinearly related parameters, are typically found by least-squares procedure derived by (30). A similar recent problem is the characterization of planets orbiting other stars from sparse time series data. Figure 1 shows a model of three planets orbiting a nearby star similar to the Sun. This model, however, may not be unique. Sophisticated Bayesian model selection techniques based on Markov chain Monte Carlo calculations are now being applied to this important problem (31).

PLACE FIGURE 1 HERE

## FAINT SOURCES AND POISSON PROCESSES

Certain fields of astronomy are devoted to the detection of individual particles or photons of light: cosmic ray and neutrino astronomy; X-ray and gamma-ray astronomy (33). The data are received as a sequence of events each with a position in the sky, energy and arrival time. Ancillary information is available from knowledge of the telescope and detectors. The goals are to identify and locate individual astronomical sources, to characterize their flux (intensity), spectrum and variability in the bands accessible to the instruments. An uninteresting background flux is present, either from noise in the detector or from cosmic emission which cannot be resolved into individual points. Telescopes in these fields are very expensive and frequently are launched into space to avoid absorption or emission from the Earth's atmosphere. Examples include the space-bourne Chandra X-ray Observatory and INTEGRAL gamma-ray observatory, the Pierre Auger Observatory in Argentina for cosmic rays, and the IceCube experiment in Antarctica for neutrinos.

Statistically, the datasets emerging from these observatories can be considered tagged Poisson processes in 4-dimensions with heteroscedastic measurement errors of known variance. While strong sources are easily found, there is a challenge in detecting the faintest sources with only a few events or only barely above background levels. Methods include sliding windows, wavelet and adaptive filtering (7, 8, 34). Variable sources are identified by the Kolmogorov-Smirnov test and are can be modeled as a sequence of constant segments derived from a Bayesian likelihood analysis (35).

When a source is not evidently present above the background, astronomers seek an upper limit to the flux that can then be considered a left-censored data point in population studies. The Poisson upper limit problem has been widely discussed and no consensus has emerged on a best method (36). A common procedure used in the particle physics community where similar problems are encountered is the Feldman-Cousins maximum likelihood limit (37). A related matter involves the ratio of two Poisson-distributed variables; this is often called the hardness ratio when the quantities are the counts in two spectral bands for a single source. Here the maximum likelihood solution is unstable (38). A Bayesian solution based on MCMC computations has been developed when background counts must be subtracted from both quantities (39).

## GALAXY CLUSTERING AND SPATIAL POINT PROCESSES

Over the past century, several lines of astronomical inquiry powerfully indicate that the Universe began in a Big Bang explosion about 14 billion years ago (40). The evidence includes: the redshifts of galaxies proportional to their distance from us (Hubble's Law); the blackbody cosmic microwave background (CMB) radiation from the early radiation-dominated era; the cosmic abundances of helium and other light element isotopes; and the statistical distribution of galaxies in space.

The galaxy distribution can be viewed as a spatial point process in three dimensions, two from location in the sky and one from redshift as a surrogate for distance. Galaxy redshift surveys reveal remarkable nonlinear, anisotropic clustering. They are preferentially located along curved sheets surrounding voids in a pattern resembling a collection of soap bubbles (Figure 2 top, 41).

The most common statistical analyses are based on the spatial Fourier spectrum and the spatial two-point correlation function that is essentially the differential of Ripley's K function for point processes (9).  While these simple functionalities do not capture the anisotropies of the distribution, they can constrain cosmological theories: Figure 2 (bottom, 42) shows the discovery of a feature predicted from a physical phenomenon in the early Universe known as Baryonic Acoustic Oscillations.

PLACE FIGURE 2 HERE

## COSMIC MICROWAVE BACKGROUND AND MODEL SELECTION

Statistical analysis of the CMB, a diffuse light with spectrum peaking around 1 millimeter wavelength, has provided a wealth of insights into modern quantitative cosmology.  Initially found to be very smooth across the sky (except for irrelevant emission from our Milky Way Galaxy), spatial fluctuations with amplitudes around a few parts per million were recently discovered.  These were predicted by theories of the Big Bang, and detailed studies of CMB fluctuations strongly constrain cosmological models.  Figure 3 shows a map of CMB spatial variations across the sky and the best-fit model to the amplitude of fluctuations as a function of spatial scale.  The model indicates that the principal constituents of the Universe are approximately: 5% ordinary matter, 23% attractive Dark Matter of unknown nature, 72% repulsive Dark Energy of unknown nature, and small contributions by photons of light and neutrinos.

A wide range of Big Bang cosmological models, both nested and non-nested, with a dozen or more parameters can be compared to the CMB fluctuation spectrum.   Astronomers seek both to exclude incorrect models and to establish parameter values and confidence intervals for permitted models.  Best-fit models have been calculated using maximum likelihood with confidence intervals derived from the Fisher Information Matrix (43, 44), nonparametric regression (45), and a Bayesian fusion of disparate observations (46).   Bayesian analyses play an increasingly important role in the interpretation of cosmological data (47).

PLACE FIGURE 3 HERE

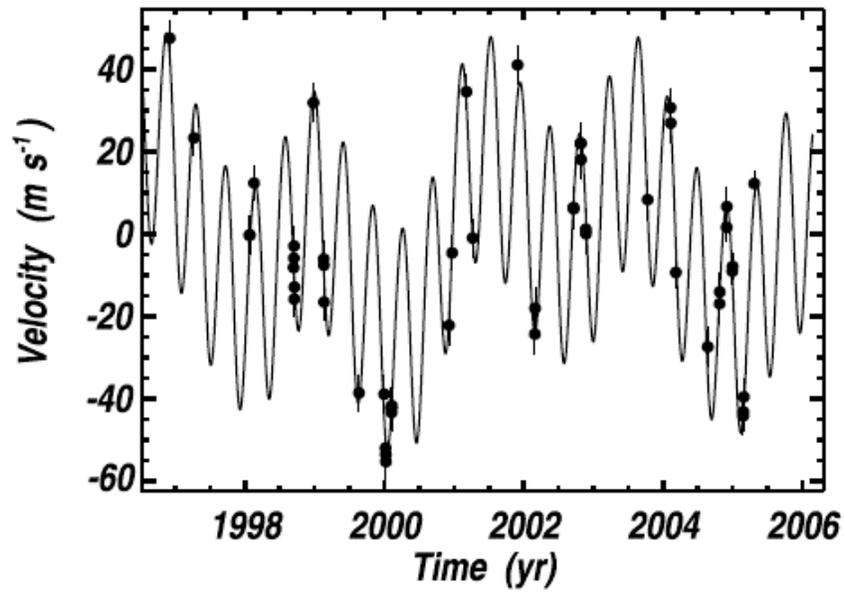

Figure 1 [ESS_Vogt.tif]  Parametric modeling of the motions of a nearby star due to the influence of orbiting planets (32).  Filled circles show the unevenly-spaced time series of radial velocities of HD 37124.  Curve shows a model involving three planets with periods 5.7, 2.4 and 0.4 years.  (Courtesy of S. Vogt)

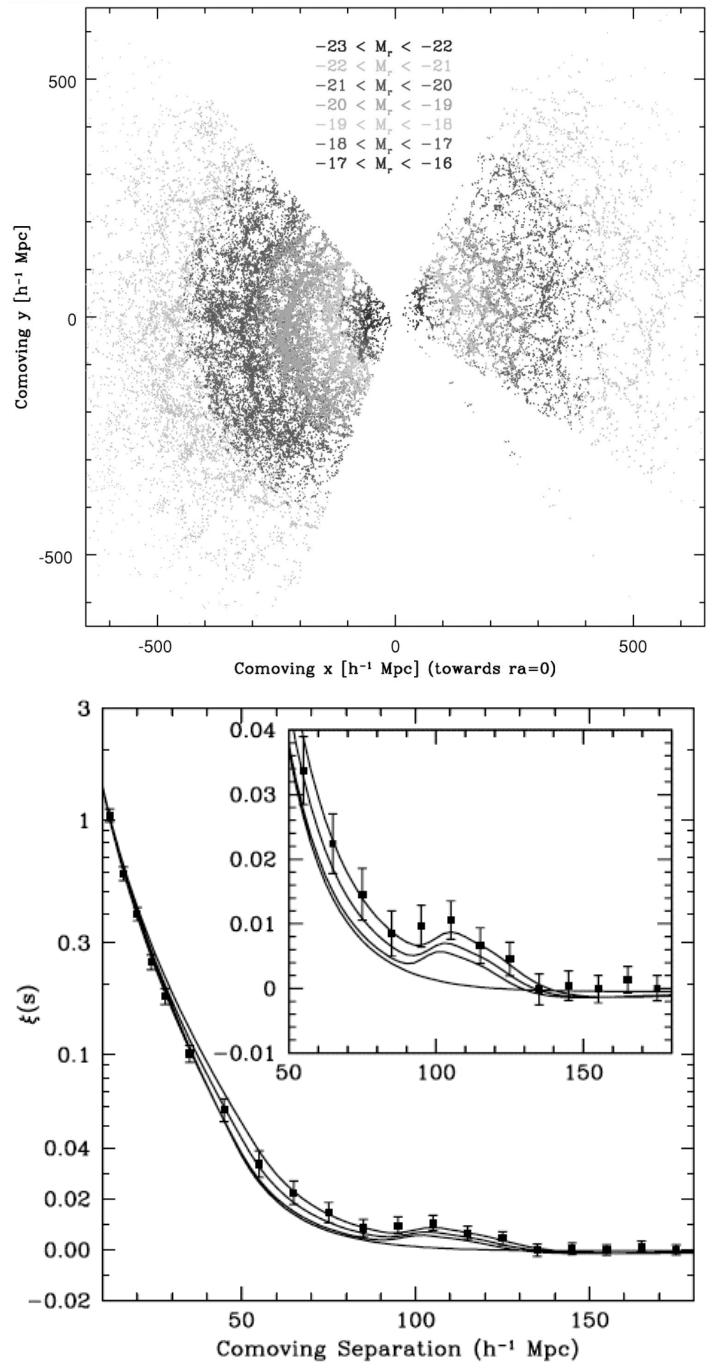

Figure 2 [Tegmark04_SDSS_gal.jpg, ESS_SDSS_BAO.tif] Galaxy clustering derived from the Sloan Digital Sky Survey. Top: Distribution of 67,676 galaxies in two slices of the sky showing strong anisotropic clustering. Bottom: The spatial two-point correlation function showing the faint feature around 100 megaparsec scales revealing cosmological Baryonic Acoustic Oscillations. (Courtesy of M. Tegmark and D. Eisenstein)

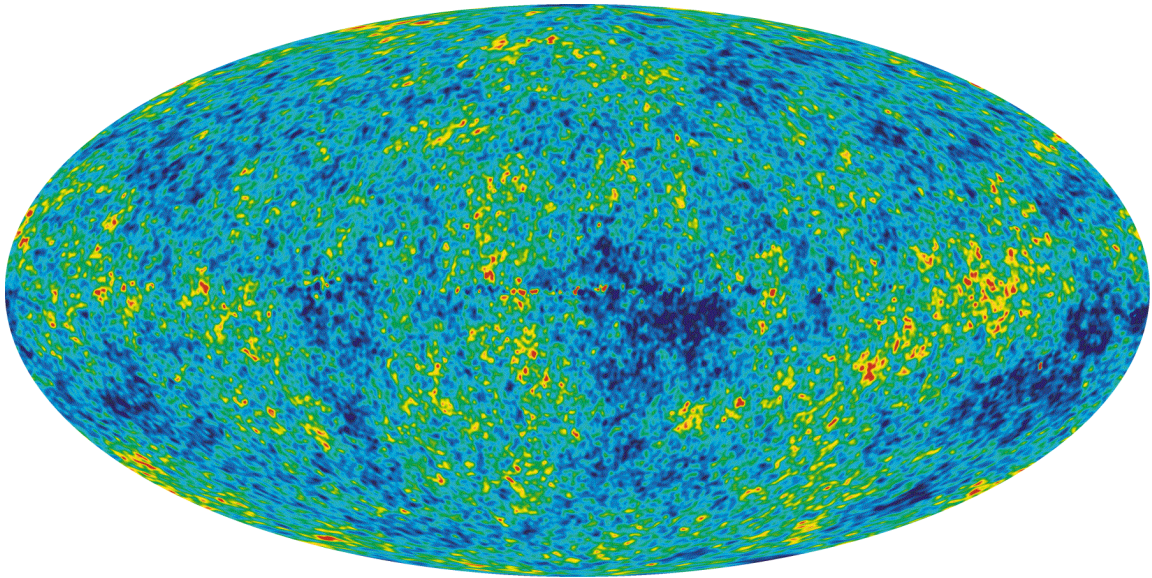

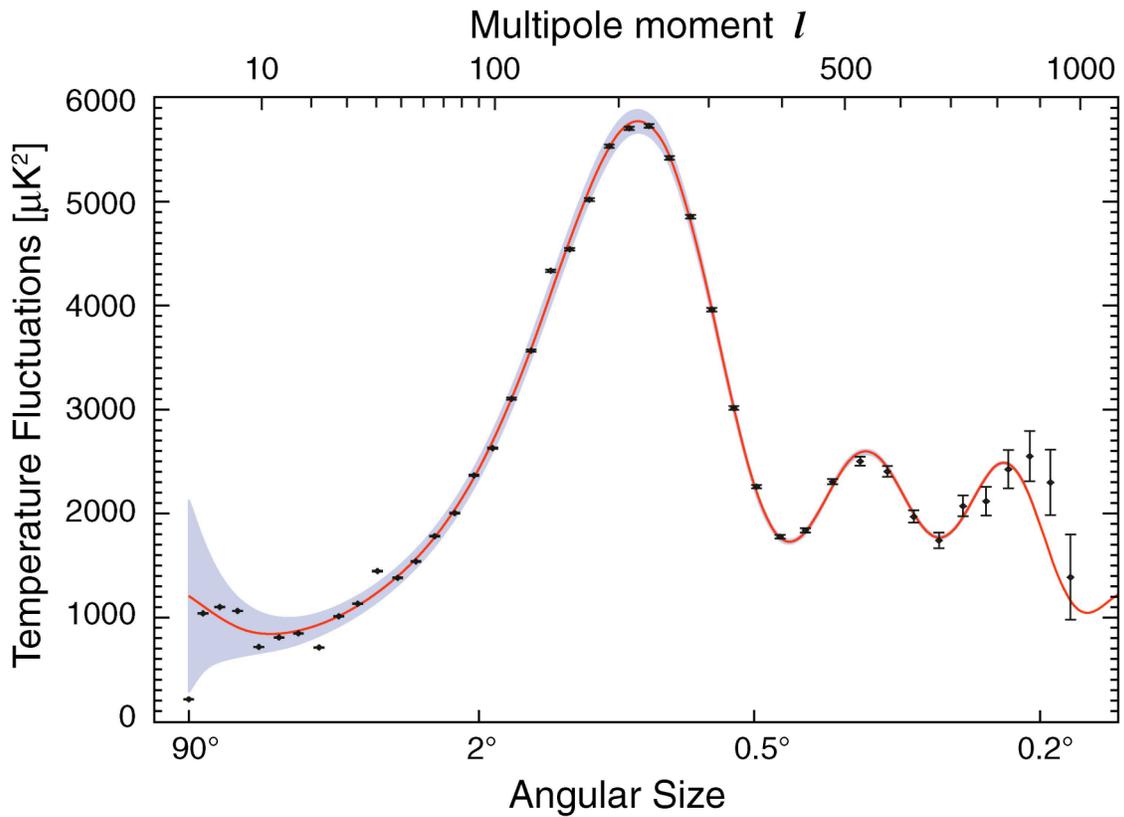

Figure 3 [08997_5yrFullSky_WMAP_4096W.tif, 080999_PowerSpectrumL.tif] Fluctuations of the cosmic microwave background radiation. Top: All-sky map produced by the WMAP satellite. Bottom: Power of fluctuations as a function of spatial scale derived from the WMAP map with the consensus Big Bang cosmological model dominated by Dark Energy and Dark Matter. (Courtesy of NASA/WMAP Science Team)